\documentclass[final,authoryear,5p,times,twocolumn]{elsarticle}

\usepackage{graphics}
\graphicspath{{fig/}}
\DeclareGraphicsExtensions{.pdf,.jpeg,.png,eps}

\usepackage{mdwmath}
\usepackage{mdwtab}
\usepackage{rotating}
\usepackage{stfloats}
\usepackage[font=small]{caption}
\usepackage[justification=centering]{caption}
\usepackage{multirow}
\usepackage{amssymb}
\usepackage[cmex10]{amsmath}
\usepackage[tight,footnotesize]{subfigure}
\usepackage[font=footnotesize]{subfig}
\usepackage{booktabs}
\usepackage{algorithm}
\usepackage{algorithmic}
\usepackage{xcolor}
\usepackage{float}
\usepackage{boxedminipage}
\usepackage[authoryear]{natbib}
\emergencystretch=\maxdimen
\biboptions{numbers,sort&compress}

\begin{document}

\begin{frontmatter}



\title{Simplification of Training Data for Cross-Project Defect Prediction}


\author[a,b]{Peng He}
\author[c,d]{Bing Li}
\author[a,b]{Deguang Zhang}
\author[b,d]{Yutao Ma \corref{cor1}}

\address[a]{State Key Lab of Software Engineering, Wuhan University, Wuhan 430072, China}
\address[b]{School of Computer, Wuhan University, Wuhan 430072, China}
\address[c]{International School of Software, Wuhan University, Wuhan 430079, China}
\address[d]{Research Center for Complex Network, Wuhan University, Wuhan 430072, China}
\cortext[cor1]{Corresponding author. Tel: +86 27 68776081 \\   E-mail: \{penghe (P. He), bingli (B. Li), deguangzhang (D.G. Zhang), ytma (Y.T. Ma)\}@whu.edu.cn}

\begin{abstract}

  Cross-project defect prediction (CPDP) plays an important role in estimating the most likely defect-prone software components, especially for new or inactive projects. To the best of our knowledge, few prior studies provide explicit guidelines on how to select suitable training data of quality from a large number of public software repositories. In this paper, we have proposed a training data simplification method for practical CPDP in consideration of multiple levels of granularity and filtering strategies for data sets. In addition, we have also provided quantitative evidence on the selection of a suitable filter in terms of defect-proneness ratio. Based on an empirical study on 34 releases of 10 open-source projects, we have elaborately compared the prediction performance of different defect predictors built with five well-known classifiers using training data simplified at different levels of granularity and with two popular filters. The results indicate that when using the multi-granularity simplification method with an appropriate filter, the prediction models based on Na\"{\i}ve Bayes can achieve fairly good performance and outperform the benchmark method.

\end{abstract}


\end{frontmatter}


\section{Introduction}
Software defect prediction is a research field that seeks effective methods for predicting the defect-proneness in a given software component. These methods can help software engineers allocate limited resources to those components that are most likely to contain defects in testing and maintenance activities. Early studies in this field usually focused on Within-Project Defect Prediction (WPDP), which trained defect predictors from the data of historical releases in the same project and predicted defects in the upcoming releases or reported the results of cross-validation on the same data set \citep{He:An}. \cite{Zimmermann:Cross} stated that defect prediction performs well within projects as long as there is a sufficient amount of data available to train prediction models. However, such an assumption does not always hold in practice, especially for newly-created or inactive software projects. For example, \cite{Rainer: Evaluating} conducted an in-depth analysis on SourceForge\footnote{http://sourceforge.net} and found that only $1\%$ of software projects on SourceForge were actually active in terms of their metrics.

Fortunately, there are many on-line public defect data sets from other projects that are freely available and can be used as training data sets (TDSs), such as PROMISE\footnote{http://promisedata.org} and Apache\footnote{http://www.apache.org}. Thus, some researchers have been inspired to overcome the above problem of WPDP by means of Cross-Project Defect Prediction (CPDP) \citep{He:An, Zimmermann:Cross, Peters:Better, Rahman:Recalling, Briand:Assessing, Turhan:On, Herbold:Training}. In general, CPDP is the art of using the data from other projects to predict software defects in the target project with a very small amount of local data. CPDP models have been proven to be feasible by many previous studies \citep{He:An,Rahman:Recalling}. However, \cite{He:An} found that the overall performance of CPDP was drastically improved with suitable training data, while \cite{Turhan:On} also affirmed that using a complete TDS would lead to excessive false alarms. That is, data quality, rather than the total quantity of data, is more likely to affect the outcomes of CPDP to some extent.

There is no doubt that the availability of defect data sets on the Internet will continue to grow, as will the popularity of open-source software. The construction of an appropriate TDS of quality gathered from a large number of public software repositories is still a challenge for CPDP \citep{Herbold:Training}. To the best of our knowledge, there are two primary ways to investigate this issue. On the one hand, many researchers have attempted to reduce data dimensions using feature selection techniques, and numerous studies have validated that a reduced feature subset can improve the performance and efficiency of defect prediction \citep{Lu:Software,Hep: An}. On the other hand, few researchers have attempted to simplify a TDS by reducing the volume of data \citep{He:An,Peters:Better} to exclude irrelevant training data and retain those that are most suitable.

\begin{figure*}
\centering
\includegraphics[width=6in]{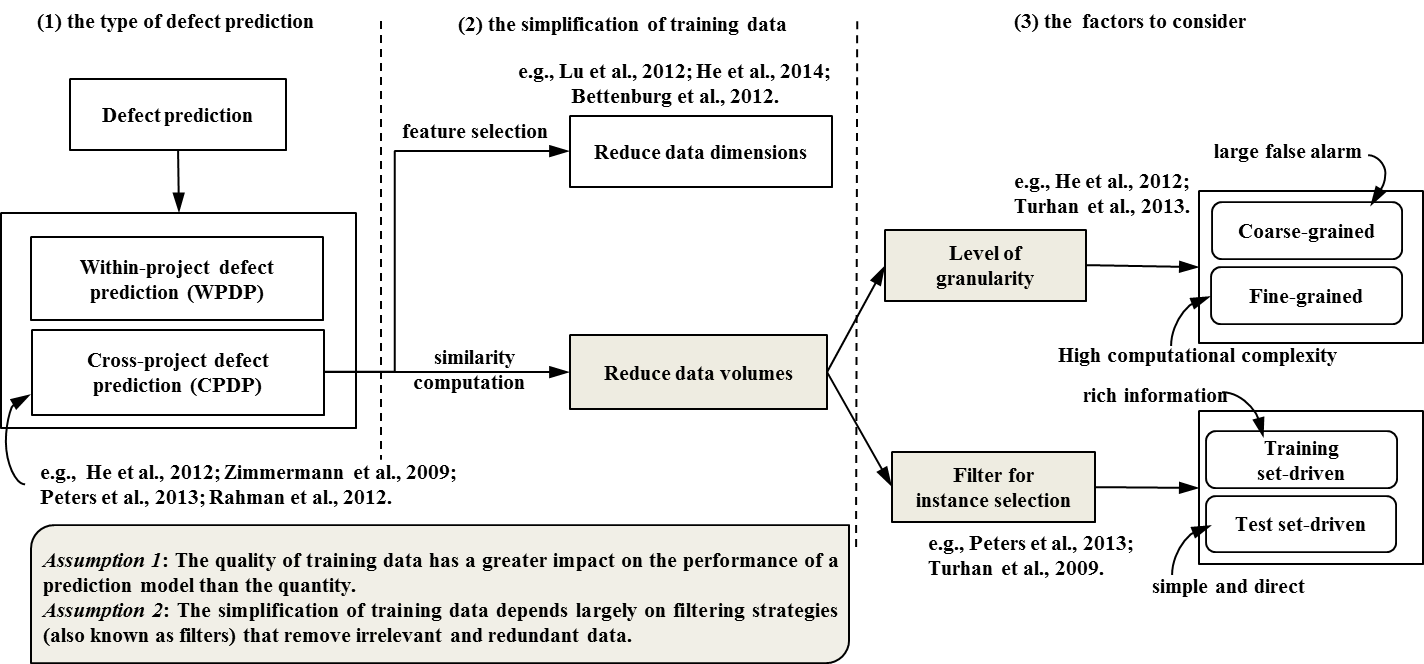}
\caption{A summary of the state-of-the-art CPDP from the perspective of training data simplification.}
\label{summary}
\end{figure*}

Figure \ref{summary} shows a simple summary of the state-of-the-art methods related to the topics of interest in this paper (see the contents with a gray background). Prior studies have attempted to reduce irrelevant training data at different levels of granularity, e.g., release-level \citep{He:An} and instance/file-level \citep{Turhan:On}. Unfortunately, they all dealt with training data simplification based on a single level of granularity. Furthermore, different filtering strategies were recently proposed to improve the selection of suitable training instances in a TDS \citep{Peters:Better}. Although these methods seem very promising separately, we actually do not know how to choose the most appropriate filter when dealing with a specific defect data set of a given project. In other words, they did not offer any practical guidelines for the decision-making on which granularity, strategy for instance selection and classifier should be preferably selected in a specific scenario.

Considering the importance of defect prediction in software development and maintenance phases, TDS simplification on data volume is the key to achieving better prediction results, as the data from other projects available on the Internet is ever-increasing. As shown in Figure \ref{summary}, to obtain an appropriate TDS of quality, we should take the two chief factors affecting training data simplification into account. Hence, the goal of this study is to propose a method to simplify a large amount of training data for CPDP in terms of different levels of granularity and filtering strategies for instance selection. We also attempt to discover useful guiding principles that can assist software engineers in building suitable defect predictors. To accomplish the above goals, we focus mainly on exploring the following research questions:
\begin{description}
  \item \emph{RQ1: Does our TDS simplification method perform well compared with the benchmark methods?} \\
  The quality of training data is one of the important factors that determine the performance of a defect predictor. TDS simplification is performed to obtain high quality training data by removing irrelevant and redundant instances. The state-of-the-art simplification methods are designed at a single level of granularity of data, and each one has its good and bad points. Hence, the goal of this research question is to examine whether our method based on a multi-granularity simplification strategy performs as well as (or outperforms) those up-to-date methods.
  \item \emph{RQ2: Which classifier is more suitable for CPDP with our TDS simplification method?}\\
  The findings of previous studies indicate that some simple classifiers perform well for CPDP without training data simplification, such as Logistic Regression and Na\"{\i}ve Bayes \citep{Hall:A}. For this research question, we would like to validate whether simple classifiers can also achieve better prediction results based on a simplified TDS.
  \item \emph{RQ3: Which filter for instance selection should be preferable in a specific scenario?}\\
  The filtering strategy (also known as the filter) determines how those appropriate instances in a TDS are selected and preserved. Currently, two types of filters for instance selection exist, i.e., training set-driven filter and test set-driven filter. However, the application contexts of the two filters remain unclear. Thus, the goal of this research question is to find a quantitative rule for filter selection, to improve the prediction performance based on a single type of filter.
\end{description}

The contribution of our work is twofold:
\begin{itemize}
  \item We proposed a multi-granularity TDS simplification method to obtain training data of quality for CPDP. Empirical results show that our method can filter out more irrelevant and redundant data compared with the benchmark method. Moreover, the predictors trained by the simplified TDS according to the method can achieve better prediction precision as a whole.
  \item We first provided practical decision rules for an appropriate choice between the two existing filtering strategies for training data simplification in terms of defect-proneness ratio. Empirical results show that the reasonable selection of filters can lead to better prediction performance than a single type of filter.
\end{itemize}

We believe that the results of our study could be a stepping stone for current and future approaches to practical CPDP, as well as a new attempt for software engineering data simplification with new learning techniques such as transfer learning in the era of Big Data.

The rest of this paper is organized as follows. Section 2 is a review of related work. In Section 3, we introduce the method for TDS simplification in detail, and in Section 4, we evaluate our experiments with a case study based on 10 open-source projects. Section 5 and Section 6 present and discuss our findings and the threats to validity, respectively. Finally, we conclude this paper and present an agenda for future work in Section 7.

\section{Related Work}

\subsection{Cross-Project Defect Prediction}
Because it is sometimes difficult for WPDP to collect sufficient historical data, CPDP is currently popular within the field of defect prediction. To the best of our knowledge, \cite{Briand:Assessing} conducted the earliest study on CPDP, and they applied the prediction model built on Xpose to Jwriter. The authors validated that such a model performed better than the random model and outperformed it in terms of class size. However, \cite{Zimmermann:Cross} conducted a large-scale experiment on data vs. domain vs. process, and found that only $3.4\%$ of 622 cross-project predictions actually worked. Interestingly, CPDP was not symmetrical between Firefox and Microsoft IE, that is, Firefox is a sound defect predictor for Microsoft IE, but not vice versa. Similar results are reported in \citep{Menzies: Local,Posnett:Ecological,Bettenburg: Think}.

\cite{Turhan:On} proposed a nearest-neighbor filtering technique to prune away irrelevant cross-project data, and they analyzed the performance of CPDP based on 10 projects collected from the PROMISE repository. Moreover, they investigated the case where prediction models were constructed from a blend of within- and cross-project data, and concluded that in case there was limited local data (e.g., $10\%$ of historical data) of a target project, such mixed project predictions were viable, as they performed as well as within-project prediction models \citep{Turhan:Empirical}.

\cite{Rahman:Recalling} conducted a cost-sensitive analysis on the efficacy of CPDP based on 38 releases of nine large Apache Software Foundation (ASF) projects. Their findings revealed that the cost-sensitive cross-project prediction performance was not worse than the within-project prediction performance, and it was substantially better than the random prediction performance. \cite{Peters:Better} introduced a new filter to realize better cross-company learning compared with the state-of-the-art Burak filter \citep{Turhan:On}. The results showed that their approach could build $64\%$ more useful predictors than both within-company and cross-company approaches based on the Burak filter, and demonstrated that the training set-driven filter was able to achieve better prediction results for those projects without sufficient local data.

\cite{He:An} conducted three experiments on the same data sets used in this study to test and verify the idea that training data from other projects can provide acceptable prediction results. They further proposed an approach to automatically select suitable training data for those projects that lack local historical data. Towards efficient training data selection for CPDP, \cite{Herbold:Training} proposed several useful strategies according to 44 data sets from 14 open-source projects. The results demonstrated that their selection strategies improved the achieved success rate of CPDP significantly, but the quality of the results was still unable to outstrip WPDP.

The review reveals that previous studies focused mainly on the feasibility of CPDP and the selection of suitable training data at a single level of granularity of data. However, relatively little attention has been paid to empirically exploring the impact of  TDS simplification in terms of different levels of granularity on prediction performance. Moreover, little is known about the decision rule for a proper choice among the existing filters for instance selection.
\subsection{Defect Prediction with Transfer Learning}
Transfer learning techniques have attracted more and more attention in machine learning and data mining over the last several years \citep{Pan: A}, and the successful applications include software effort estimation \citep{Kocagune: Transfer}, text classification \citep{Xue: Topic}, name-entity recognition \citep{Arnold: A}, natural language processing \citep{Pan: Cross} and email spam filtering \citep{Zhang: A}. Recently, CPDP was also deemed as a transfer learning problem. The problem setting of CPDP is related to the adaptation setting in transfer learning for building a classifier in the target project using the training data from those relevant source projects. Thus far, transfer learning techniques have been proven to be appropriate for CPDP in practice \citep{Nam:Transfer}.

To harness cross-company defect datasets, \cite{Ma: Transfer} utilized the transfer learning method to build faster and highly effective prediction models. They proposed a novel algorithm that used the information of all the suitable features in training data, known as Transfer Na\"{\i}ve Bayes (TNB), and the experimental result indicated that TNB was more accurate in terms of AUC (the area under the receiver operating characteristic curve) and less time-consuming than benchmark methods.

\cite{Nam:Transfer} applied the transfer learning method, called TCA (Transfer Component Analysis), to find a latent feature space for the data of both training and test projects by minimizing the distance between the data distributions while preserving the original data properties. After learning the latent space in terms of six statistical characteristics, i.e., \emph{mean, median, min, max, standard deviation} and the number of instances, the data of training and test projects will be mapped onto it to reduce the difference in the data distributions. The experimental results for eight open-source projects indicated that their method significantly improved CPDP performance.

In general, although the above studies improve the performance of CPDP, they are time-consuming in that their experiments were conducted at the level of instances (files). In this study, to overcome the data distribution difference between source and target projects, we have also adopted the transfer learning method, which was applied to the releases available from different projects.

%

%

\begin{figure*}
\centering
\includegraphics[width=6in,height=1.2in]{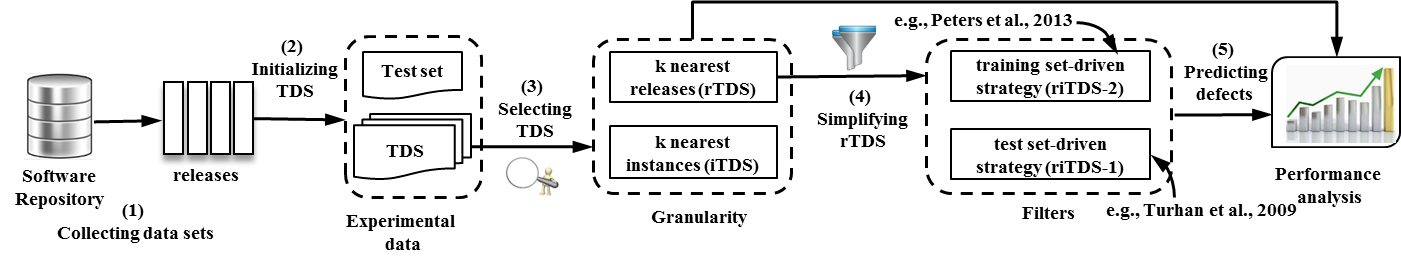}
\caption{The process of TDS simplification for CPDP.}
\label{process}
\end{figure*}

\section{Methodology}
In this paper, CPDP is defined as follows: Given a source project $P_{S}$ and a target project $P_{T}$, CPDP aims to achieve the target prediction in $P_{T}$ using the knowledge extracted from $P_{S}$, where $P_{T}\neq P_{S}$. Assuming that source and target projects have the same set of features, they may differ in feature distribution characteristics. The goal of our method is to learn a model from the selected source projects (training data) and apply the learned model to a target project (test data). Based on prior studies on CPDP, the TDS simplification process for CPDP is both explained in the following paragraphs and illustrated in Figure \ref{process}. Specifically, unlike previous studies, we introduce two levels of granularity and two types of filtering strategies for TDS simplification based on characteristic and instance vectors.

In brief, our method for TDS simplification has two key steps. The first step is selecting $k$ candidate releases that are most similar to the target release in terms of data distributional characteristics. The second is choosing the $k$ nearest instances of each test instance from those candidate releases according to suitable filtering strategies. Based on different classifiers, defect predictors can be trained from the simplified TDS, and then are applied to test data.

\begin{figure}
\centering
\includegraphics[width=3in,height=1.8in]{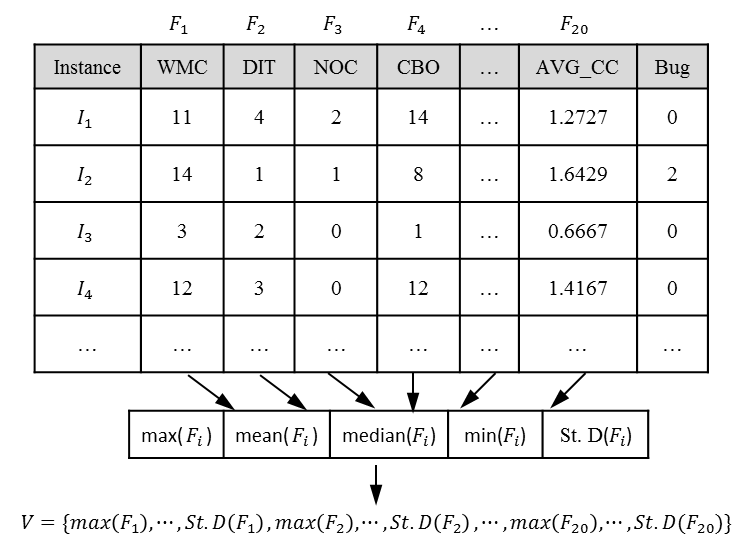}
\caption{The structure of a release ($R$) (instances ($I$), features ($F$) and distributional characteristics ($V$)): an example.}
\label{release}
\end{figure}

In our context, a release $R$ contains $m$ instances (.java files), represented as $R=\{I_{1},I_{2},\cdots,I_{m}\}$. An instance can be represented as $I_{i}=\{f_{i1}, f_{i2},\cdots,f_{in}\}$, where $f_{ij}$ is the $j^{th}$ feature value of the instance $I_{i}$, and $n$ is the number of features. Meanwhile, a feature vector can be represented as $F_{i}=\{f_{1i}, f_{2i},\cdots,f_{mi}\}$, where $f_{ji}$ is the value of the $j^{th}$ instance for the feature $F_{i}$, and $m$ is the number of instances.  An initial TDS\textemdash an aggregate of multiple data sets\textemdash is often comprised of many releases from different projects: $S=\{R_1,R_2,\cdots,R_l\}$, where $l$ is the number of releases. The distributional characteristic vector of a release can be formulated as $V=\{C_1,C_2,\cdots,C_k,\cdots,C_n\}$, where $C_k$ is the distribution of the feature $F_{k}$ and can be written as $C_{k}=\{SC_{1},SC_{2},\cdots,SC_{s}\}$ (see Figure \ref{release}). For the meaning of the statistical characteristics $SC_s$, please refer to Table \ref{characteristic}.

\subsection{Level of Granularity}

For CPDP, one of the easiest methods is to directly train prediction models without any TDS simplification methods. During this learning process, all of the data from other projects are utilized as a TDS. Take the experimental datasets used in this paper as an example; Table \ref{no} shows the prediction results of CPDP without TDS simplification. Clearly, the average number of training instances is much greater than the size of each test set. More detailed information of the experimental datasets will be introduced in Section 4.1. In fact, our experimental data occupied a very small fraction of the public defect data available on the Internet. On the one hand, although it does not matter for computing resources and time complexity, for a learning process based on a vast amount of training data, it is cost-sensitive and not practical for software engineers; on the other hand, this decreases the accuracy of prediction models to some extent \citep{Turhan:On,He:An}. Therefore, how to obtain the right training data by TDS simplification becomes meaningful \citep{Peters:Better}.

\begin{table}\small
\centering
\caption{Description of the indicators used to describe the distributional characteristics of a release}\label{characteristic}
\begin{tabular}[c]{p{12mm}p{70mm}}
  \hline
  Indicator & Description \\  \hline
  Median  & The numeric value separating the higher half of a population from the lower half   \\
  Mean  & The average value of samples in a population; specifically, it refers to arithmetic mean in this paper \\
  Min  & The least value in a population \\
  Max  & The greatest value in a population \\
  St. D  & The square root of the variance \\
  \hline
\end{tabular}
\end{table}

\textbf{rTDS:} The TDS simplification at the release level is a simple and coarse-grained method, referred to as \emph{rTDS}. The coarse-grained simplification of training data often uses the k-Nearest Neighbors algorithm to measure the similarity (via Euclidean distance\footnote{http://en.wikipedia.org/wiki/Euclidean$\_$ distance}) between the release $V_{training}$ and the release $V_{target}$. That is, the $k$ nearest candidate releases are selected as the ultimate TDS \citep{He:An,Turhan:On,Herbold:Training}. In our study, a data set is a release of a project, and five commonly-used indicators, i.e., \emph{max}, \emph{min}, \emph{median}, \emph{mean}, and $standard\quad deviation$, are involved in describing the statistical characteristics (SCs) of a release (see Table \ref{characteristic}). Thus, the distance between two releases can be formulated as: $distance_R=\sqrt{(SC_{i1}-SC_{j1})^2+\cdots +(SC_{is}-SC_{js})^2}$.

\begin{table}\small
\centering
\caption{ The results of CPDP without TDS simplification. Numeric values in the second and third columns indicate the mean values of the measures. \# instances (TDS) represents the average number of training instances in all TDSs in question.}\label{no}
\begin{tabular}{c|c|c|c}
  \hline
   Classifiers & f-measure & g-measure & \# instances (TDS) \\ \hline
   J48 & 0.369 & 0.499 & \multirow{5}*{11824} \\
   LR  & 0.291 & 0.358 &   \\
   NB  & 0.464 & 0.617 &   \\
   RF  & 0.322 & 0.432 &   \\
   SVM & 0.311 & 0.392 &   \\
  \hline
\end{tabular}
\end{table}

\textbf{iTDS:} Compared with the \emph{rTDS}, the fine-grained TDS simplification should be conducted based on the computation of the similarity between the instance $I_{training}$ and the instance $I_{target}$, which is referred to as \emph{iTDS}. It returns the $k$ nearest training instances for each target instance $I_{target}$ by calculating their Euclidean distance \citep{Peters:Better,Nam:Transfer}. Thus, the distance between two instances can be formulated as: $distance_I=\sqrt{(f_{i1}-f_{j1})^2+\cdots +(f_{in}-f_{jn})^2}$.


\textbf{riTDS:} Considering that there are a large number of on-line public defect data sets available for use as candidate training data, and that the number is still growing fast, it is impractical to completely calculate the distances of all instance pairs by the \emph{iTDS}. However, using the \emph{rTDS} alone may cause excessive false alarms because of the inclusion of many irrelevant training instances. Thus, we propose a two-step strategy for TDS simplification\textemdash \emph{riTDS}, which obtains the coarse-grained set rTDS first and then simplifies it by a fine-grained method such as the \emph{iTDS}. This strategy can be interpreted as a combination of the aforementioned two cases, also named as a multi-granularity simplification strategy. In other words, we first select the $k$ nearest releases, instead of all releases available, as the candidate training data rTDS. Subsequently, we further simplify the coarse-grained set rTDS at the instance level according to suitable filters.


\subsection{Filter for Instance Selection}
For the \emph{riTDS}, in the second step, there are two state-of-the-art filters for instance selection according to the choice of reference data. One is driven by the test set and returns the $k$ nearest instances in the set rTDS for each test instance directly (abbreviated to \emph{riTDS-1} in our context). This filter is to ensure that the information of each test instance is fully utilized, and it is referred to as a test set-driven filter. The other is just the opposite; it is training data-driven via labeling of the $k$ nearest test instances for each training instance first and then returning of the nearest training instance of each labeled test instance (abbreviated to \emph{riTDS-2} in our context). Clearly, in this case, it is possible that some test instances are never labeled as the nearest instance for certain training instances. Therefore, not all test instances will be utilized in favor of training instances.
\begin{figure}[H]
\centering
\includegraphics[width=3in,height=2in]{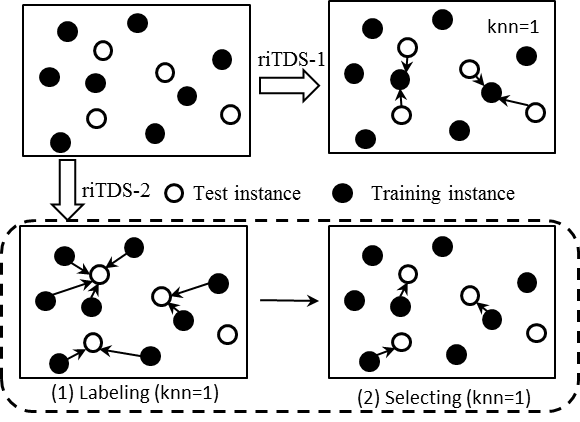}
\caption{The description of two types of filters for instance selection.}
\label{riTDS}
\end{figure}

The informal description of these two types of filters is shown in Figure \ref{riTDS}.  To the best of our knowledge, the Burak filter \citep{Turhan:On} and the Peters filter \citep{Peters:Better} are the typical representatives of these two types of filters. For more details of their implementation, please refer to the related literature. Note that our primary goal in this section is to find some helpful guidelines for software engineers to definitely discriminate the application contexts of each filter, instead of improving the performance of these existing filters. Algorithm \ref{Algorithm1} formalizes the implementation of the \emph{riTDS} with regard to these two filters.

\section{Case Study}
\subsection{Data Setup}
 In this study, 34 releases of 10 open-source projects available in the PROMISE repository are used for our experiments. Detailed information of the releases is listed in Table \ref{data}, where $\#$Instances represents the number of instances in a release, and the number of defects and the proportion of buggy instances are listed in the corresponding columns $\#$Defects and $\%$Defects, respectively. Each instance in a release represents a class ($.java$) file and consists of 20 software metrics (independent variables) and a binary label for the defect proneness (dependent variable). Table \ref{metrics} presents all metrics used in this study as well as their descriptions. For those readers who are interested in the datasets, please refer to \citep{Jureczko:Towards}.

Before performing a cross-project defect prediction, we need to select a target data set and its appropriate TDS. Each one in the 34 releases was selected to be the target data set once, i.e., we repeated our approach for 34 different cross-project defect predictions. With regard to our primary objective, we set up an initial TDS for CPDP, which excluded any releases from the target project. For instance, for Xalan-2.5, the releases Xalan-2.4 and Xalan-2.6 cannot be included in its initial TDS.

\begin{algorithm}[H]\small
\renewcommand{\algorithmicrequire}{\textbf{Input:}}
\renewcommand{\algorithmicensure}{\textbf{Method:}}
  \caption{A two-step strategy for TDS simplification}\label{Algorithm1}
 \begin{algorithmic}[1]
 \REQUIRE
   \STATE Candidate TDS set  $S=\{R_{1},R_{2},\cdots,R_{N}\}$; \
   \STATE Target release $R_{target}=\{I_{1},I_{2},\cdots,I_{m}\}$; \
   \STATE Number of selected releases $r$;
   \STATE Filtering strategy $F=\{$\emph{training set-driven}, \emph{test set-driven}$\}$;
 \ENSURE
   \STATE Let $rTDS$ be the top $r$ nearest releases of $R_{target}$ in $S$;
   \STATE Let $riTDS$ be the simplified training set;
   \STATE Initialize $rTDS \leftarrow \varnothing$, $riTDS\leftarrow \varnothing$, $r=3$;
   \WHILE {$r > 0$}
      \STATE  // $r=1,2,3$ in this paper
      \STATE // return the $r$ nearest releases for $R_{target}$ in terms of $distance_R$
      \STATE $rTDS\leftarrow KNN(S, R_{target}, r)$;
      \IF {$F\leftarrow$ \emph{training set-driven}}
         \FOR {each instance $I \in R_{target}$}
            \STATE // return its $k$ nearest instances in $rTDS$ in terms of $distance_I$
            \STATE $tempSet\leftarrow KNN(rTDS, I, k)$;
          \ENDFOR
          \STATE $riTDS\leftarrow tempSet$;
      \ELSE
         \FOR {each instance $I \in rTDS$}
            \STATE // label its $k$ nearest instances in $R_{target}$ in terms of $distance_I$
            \STATE $labelMap\leftarrow Label(I,R_{target},k)$;
            \STATE $tempSet\leftarrow$ the set of labeled target instances;
         \ENDFOR
         \FOR {each instance $I \in tempSet$}
            \STATE // return its nearest instance $I' (I' \in rTDS)$ according \\
            to the $labelMap$, if a test instance's nearest instance \\
            has been chosen, select the next nearest one.
            \STATE $riTDS\leftarrow riTDS \cup \{I'\}$;
         \ENDFOR
      \ENDIF
      \STATE $r--$;
     \ENDWHILE
  \RETURN $riTDS$;
 \end{algorithmic}
\end{algorithm}

Note that, there is a preprocessing that transforms the bug attribute into a binary value before using it as the dependent variable in our context. According to our prior work \citep{Hep: An}, we find that the majority of class files in the 34 data sets have no more than 3 defects, and the ratio of instances with more than 10 defects to the total instances is less than 0.2\%. In a word, a class is non-buggy only if the number of bugs in it is equal to 0. Otherwise, it is buggy regardless of the number of bugs. Similar preprocessing has been used in several prior studies, such as \citep{Peters:Better,He:An,Turhan:On,Turhan:Empirical,Herbold:Training}.

Moreover, some prior studies have suggested that a logarithmic filter on numeric values might improve prediction performance because of the highly skewed distribution of feature values \citep{Menzies: Metrics,Turhan:On}. In this paper, for each numeric value $f_{ij}$, $f^{'}_{ij}=ln(f_{ij}+1)$, where $f^{'}_{ij}$ is the new value of the original value $f_{ij}$. There are some other commonly used methods for numeric values preprocessing, such as max-min and z-score methods \citep{Nam:Transfer}.

\begin{table}[H]\small
\centering
\caption{Details of the 34 data sets, including the number of instances (files) and defects and the defect rate.}\label{data}
\begin{tabular}{c|c|c|c|c}
  \hline
  No. & Releases  & \#Instances &\#Defects &\%Defects \\ \hline
  1  & Ant-1.3   &  125   &  20   & 16.0    \\
  2  & Ant-1.4   &  178   &  40   & 22.5 \\
  3  & Ant-1.5   &  293   &  32   & 10.9 \\
  4  & Ant-1.6   &  351   &  92   & 26.2 \\
  5  & Ant-1.7   &  745   &  166  & 22.3 \\
  6  & Camel-1.0 &  339   &  13   &  3.8   \\
  7  & Camel-1.2 &  608   &  216  &  35.5 \\
  8  & Camel-1.4 &  872   &  145  & 16.6 \\
  9  & Camel-1.6 &  965   &  188  &  19.5 \\
  10 & Ivy-1.1   &  111   &  63   &  56.8 \\
  11 & Ivy-1.4   &  241   &  16   &  6.6 \\
  12 & Ivy-2.0   &  352   &  40   &  11.4 \\
  13 & Jedit-3.2 &  272   &  90   &  33.1 \\
  14 & Jedit-4.0 &  306   &  75   &  24.5 \\
  15 & Lucene-2.0&  195   &  91   &  46.7 \\
  16 & Lucene-2.2&  247   &  144  &  58.3 \\
  17 & Lucene-2.4&  340   &  203  &  59.7 \\
  18 & Poi-1.5   &  237   &  141  &  59.5 \\
  19 & Poi-2.0   &  314   &  37   &  11.8 \\
  20 & Poi-2.5   &  385   &  248  &  64.4 \\
  21 & Poi-3.0   &  442   &  281  &  63.6 \\
  22 & Synapse-1.0 &  157   &  16   &  10.2 \\
  23 & Synapse-1.1 &  222   &  60   &  27.0 \\
  24 & Synapse-1.2 &  256   &  86   &  33.6 \\
  25 & Velocity-1.4&  196   &  147  &  75.0 \\
  26 & Velocity-1.5&  214   &  142  &  66.4 \\
  27 & Velocity-1.6&  229   &  78   &  34.1 \\
  28 & Xalan-2.4   &  723   &  110  &  15.2 \\
  29 & Xalan-2.5   &  803   &  387  &  48.2 \\
  30 & Xalan-2.6   &  885   &  411  &  46.4 \\
  31 & Xerces-init &  162   &  77   &  47.5 \\
  32 & Xerces-1.2  &  440   &  71   &  16.1 \\
  33 & Xerces-1.3  &  453   &  69   &  15.2 \\
  34 & Xerces-1.4  &  588   &  437  &  74.3 \\
  \hline
\end{tabular}
\end{table}

\subsection{Experimental Design}

\begin{figure*}
\centering
\includegraphics[width=5.5in,height=2.6in]{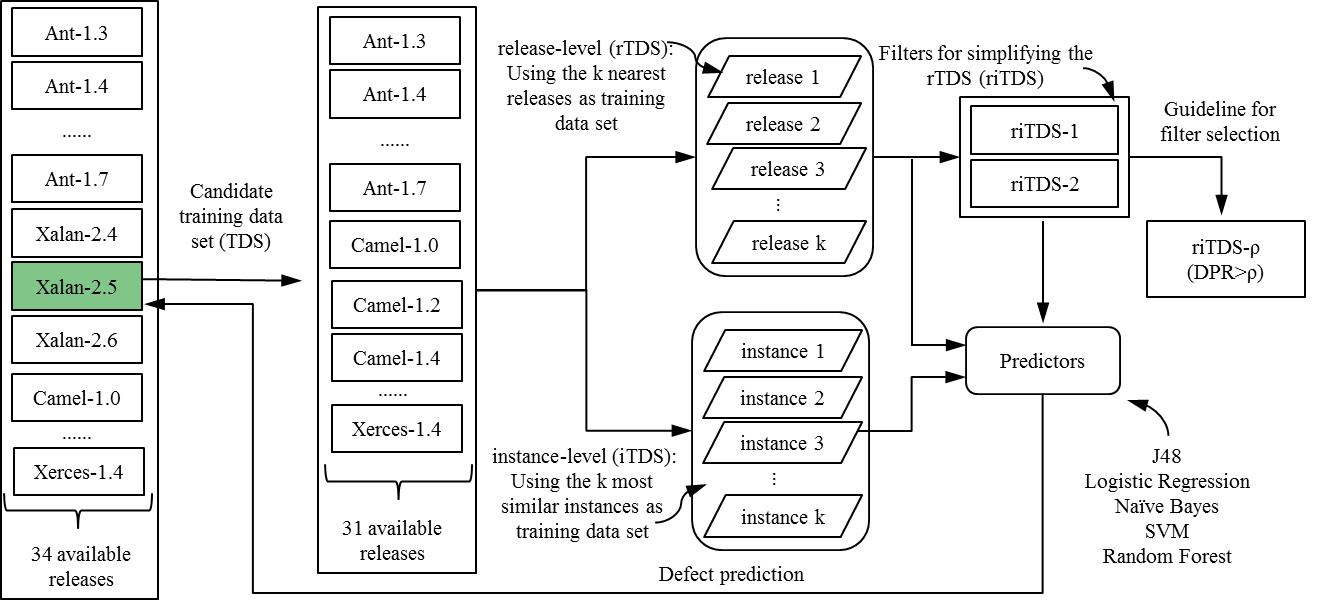}
\caption{The framework of our approach\textemdash an example of the target project Xalan-2.5.}
\label{Fig.framework}
\end{figure*}

Based on the prediction results of the predictors trained without TDS simplification (see Table \ref{no}), the entire framework of our experiments is illustrated in  Figure \ref{Fig.framework}.

First, to make a comparison between our method and the benchmark methods, three types of TDS simplification methods were considered in our experiments: (1) coarse-grained  TDS simplification (\emph{rTDS}), which uses the nearest $k$ training releases of the target release as training data; (2) fine-grained TDS simplification (\emph{iTDS}), which uses the nearest $k$ training instances of each target instance as training data; and (3) multi-granularity TDS simplification (\emph{riTDS}), which selects suitable training instances from the set rTDS. For the $rTDS$ and the $iTDS$, they were built based on a single level of granularity of data. For the $riTDS$, we designed two variants with the two filters (\emph{riTDS-1} and \emph{riTDS-2}) to simplify the set rTDS.

\begin{table}[H]\small
\centering
\caption{Description of the metrics included in the data sets.}\label{metrics}
\begin{tabular}{cc} \hline
 \textbf{Variable}&\textbf{Description} \\ \hline
 \multicolumn{2}{c}{CK suite (6)}  \\ \hline
  WMC & Weighted Methods per Class  \\
  DIT & Depth of Inheritance Tree\\
  LCOM & Lack of Cohesion in Methods \\
  RFC & Response for a Class\\
  CBO & Coupling between Object classes \\
  NOC & Number of Children\\ \hline
\multicolumn{2}{c}{Martin¡¯s metric (2)}  \\ \hline
  CA & Afferent Couplings \\
  CE & Efferent Couplings\\ \hline
\multicolumn{2}{c}{QMOOM  suite (5)} \\ \hline
  DAM & Data Access Metric \\
  NPM & Number of Public Methods\\
  MFA & Measure of Functional Abstraction\\
  CAM & Cohesion Among Methods \\
  MOA & Measure Of Aggregation \\ \hline
\multicolumn{2}{c}{Extended CK suite (4)} \\ \hline
    IC  & Inheritance Coupling \\
    CBM & Coupling Between Methods\\
    AMC & Average Method Complexity \\
    LCOM3 & Normalized version of LCOM\\ \hline
\multicolumn{2}{c}{McCabe's CC (2)} \\ \hline
    MAX$\_$CC & Maximum values of methods in the same class \\
  AVG$\_$CC & Mean values of methods in the same class \\ \hline
  LOC & Lines Of Code \\ \hline
  Bug & non-buggy or buggy\\ \hline
\end{tabular}
\end{table}

Second, we applied five typical classifiers for building defect predictors and compared their impacts on prediction results of the three types of TDS simplification methods in terms of evaluation measures.

Third, on the basis of the filtering strategies, we further sought the decision rule to determine an appropriate filter for a given data set and tested its effectiveness compared with the results of the above methods with a single type of filter.

\subsection{Classifiers}
In this study,  prediction models were built with five well-known classification algorithms\textemdash namely, J48, Logistic Regression (LR), Na\"{\i}ve Bayes (NB), Support Vector Machine (SVM) and Random Forest (RF)\textemdash used in prior studies. All classifiers were implemented in Weka\footnote{http://www.cs.waikato.ac.nz/ml/weka/}. For our experiments, we used the default parameter settings for different classifiers specified in Weka unless otherwise specified.

\textbf{J48} is an open source Java implementation of the C4.5 decision tree algorithm in Weka, which is an extension of the ID3 algorithm and uses a divide and conquer approach to growing decision trees. For each variable $X=\{x_1, x_2, \cdots, x_n\}$ and the corresponding class $Y=\{y_1, y_2,\cdots, y_m\}$, the information entropy and information gain are calculated as follows \citep{Bhargava: Decision}:

\begin{equation}\label{Entropy}
        Entropy(X)=-\sum^n_{j=1}P(x_i)logP(x_i),
\end{equation}

\begin{equation}\label{conEntropy}
        Entropy(X|Y)=\sum_{i,j}P(x_i, y_j)log\frac{P(y_j)}{P(x_i, y_j)},
\end{equation}

\begin{equation}\label{Gain}
        Gain(X,Y)=Entropy(X)-Entropy(X|Y),
\end{equation}

where $P(x_i)$ is the probability that $X=x_i$, and $P(x_i, y_j)$ is the probability that $X=x_i$ and $Y=y_j$.

\textbf{Na\"{\i}ve Bayes (NB)} is one of the simplest classifiers based on conditional probability, and it is termed as ``na\"{\i}ve" because it assumes that features are independent, that is, $P(\mathbf{X}|Y)=\prod ^n_{i=1}P(X_{i}|Y)$, where $\mathbf{X}=(X_1,\cdots, X_n)$ is a feature vector and $Y$ is a class. Although the independence assumption is often violated in the real-world, Na\"{\i}ve Bayes has been proven to be effective in many practical applications \citep{Rish: An}. A prediction model constructed by this classifier is a set of probabilities. Given a new class, the classifier estimates the probability that the class is buggy, based on the product of the individual conditional probabilities for the feature values in the class. Equation (\ref{NB}) is the fundamental equation for the Na\"{\i}ve Bayes classifier.

\begin{equation}\label{NB}
        P(Y=k|\mathbf{X})=\frac{P(Y=k)\prod_{i}P(X_{i}|Y=k)}{\sum_{j}P(Y=k)\prod_{i}P(X_{i}|Y=k)}.
\end{equation}

\textbf{Logistic Regression (LR)} is used to learn functions of the form $P(Y|\mathbf{X})$ in the case where $Y$ is a discrete value and $\mathbf{X}=( X_1, \ldots, X_n)$ is any vector containing continuous or discrete values, and it directly estimates its parameters from training data. In this paper, we will primarily consider the case where $Y$ is a binary variable (i.e., buggy or non-buggy). Note that the sum of equation (\ref{buggy}) and equation (\ref{non-buggy}) must equal 1, and $w$ is the weight \citep{Rish: An}.

\begin{equation}\label{buggy}
        P(Y=1|\mathbf{X})=\frac{1}{1+exp(w_0+\sum^n_{i=1}w_{i}X_{i})}.
\end{equation}

and
\begin{equation}\label{non-buggy}
        P(Y=0|\mathbf{X})=\frac{exp(w_0+\sum^n_{i=1}w_{i}X_{i})}{1+exp(w_0+\sum^n_{i=1}w_{i}X_{i})}.
\end{equation}

\textbf{Support Vector Machine (SVM)} is typically used for classification and regression analysis by finding the optimal hyperplane that maximally separates samples in two different classes. To classify $m$ instances in the $n-$dimensional real space $R^n$, the standard linear SVM is usually used. A prior study conducted by \cite{Lessmann: Benchmarking} showed that the SVM classifier performed as well as the Na\"{\i}ve Bayes classifier in the context of defect prediction.

\textbf{Random Forest (RF)} is a combination of tree predictors such that each tree depends on the values of a random vector sampled independently and with the same distribution for all trees in the forest \citep{Breiman: Rf}. In other words, RF is a collection of trees, where each tree is grown from a bootstrap sample. Additionally, the attributes used to find the best split at each node are a randomly chosen subset of the total number of attributes. Each tree in the collection is used to classify a new instance. The forest then selects a classification by choosing the majority result.

\subsection{Evaluation Measures}
A binary classifier can make two possible errors: \emph{false positive (FP)} and \emph{false negative (FN)}. A correctly classified defective class is a \emph{true positive (TP)} and a correctly classified non-defective class is a \emph{true negative (TN)}. The prediction performance measures used in our experiments are described as follows:


\begin{itemize}
  \item \emph{Precision (prec)} addresses how many of the defective instances returned by a model are actually defective. The higher the precision is, the fewer false positives exist.
      \begin{equation}\label{prec}
        prec =  \frac{TP}{TP+FP}.
      \end{equation}
  \item \emph{Recall (pd)} addresses how many of the defective instances are actually returned by a model. The higher the recall is, the fewer false negatives exist.
      \begin{equation}\label{pd}
        pd = \frac{TP}{TP+FN}.
      \end{equation}
  \item \emph{pf} (probability of false alarm) measures how many of the instances that triggered the predictor actually did not contain any defects. The best $pf$ value is 0.
      \begin{equation}\label{pf}
        pf=\frac{FP}{FP+TN}.
      \end{equation}
  \item \emph{f-measure} can be interpreted as a weighted average of \emph{Precision} and \emph{Recall}. The value of \emph{f-measure} ranges between 0 and 1.
      \begin{equation}\label{f-measure}
        f-measure =\frac{2*pd*prec}{pd+prec}.
      \end{equation}
  \item \emph{g-measure} (the harmonic mean of $pd$ and $1-pf$): $1-pf$ represents \emph{Specificity} (the proportion of correctly identified defect-free instances) and is used together with $pd$ to form the $G$-$mean_2$ measure. In our paper, we use these to form the \emph{g-measure} as defined in \citep{Peters:Better}.
      \begin{equation}\label{g-measure}
        g-measure =\frac{2*pd(1-pf)}{pd+(1-pf)}.
      \end{equation}
  \item \emph{Accuracy (acc)} measures how well a binary classification correctly identifies. The higher the accuracy is, the fewer errors made by a classifier exist. In this paper, it is used to measure the proportion of true recommendation results when answering the \emph{RQ3} in the following section.
      \begin{equation}\label{accuracy}
        acc= \frac{TP+TN}{TP+FP+TN+FN}.
      \end{equation}

  \item \emph{AUC} (the area under the Receiver Operating Characteristic (ROC) curve) is the portion of the area of unit square, equal to the probability that a classifier will identify a randomly chosen defective class higher than a randomly chosen defect-free one \citep{Fawcett: An}. An AUC value less than 0.5 indicates a very low true positive rate and high false alarm. As we know, compared with traditional accuracy measures, AUC is more suitable to reflect the performance of predictors regarding the problem of class distribution imbalance. Therefore, we also use AUC to evaluate the most suitable classifier for our method in \emph{RQ2}.

\end{itemize}

\begin{table*}\small
\centering
\caption{ The results of TDS simplification at different levels of granularity. The numbers in bold are the maximum among the five classifiers for each TDS simplification method in each scenario ($r=1, 2, 3$).}\label{Tab.RQ-1}
\begin{tabular}{c|c|ccc|ccc|ccc}
  \hline
  \multicolumn{1}{c|}{\multirow{2}*{ Strategies}} & \multicolumn{1}{c|}{\multirow{2}*{ Classifiers}}
  &\multicolumn{3}{c|}{f-measure} &\multicolumn{3}{c|}{g-measure} &\multicolumn{3}{c}{\#instances(simplified TDS)}\\
  \cline{3-11}
  & & \multicolumn{1}{c}{1} &\multicolumn{1}{c}{2} &\multicolumn{1}{c|}{3}
    & \multicolumn{1}{c}{1} &\multicolumn{1}{c}{2} &\multicolumn{1}{c|}{3}
    & \multicolumn{1}{c}{1} &\multicolumn{1}{c}{2} &\multicolumn{1}{c}{3}
   \\ \hline
  \multirow{5}*{\emph{rTDS}} & J48 & 0.334 & 0.348 & 0.336 & 0.402 & 0.425 & 0.425 &\multirow{5}*{387.4} & \multirow{5}*{798.1} & \multirow{5}*{1222.7} \\
   & LR  & 0.322 & 0.342 & 0.336 & 0.385 & 0.427 & 0.416 & & &  \\
   & NB  & \textbf{0.435} & \textbf{0.459} & \textbf{0.459} & \textbf{0.552} & \textbf{0.592} & \textbf{0.594} & & &  \\
   & RF  & 0.305 & 0.316 & 0.299 & 0.354 & 0.404 & 0.390 & & &  \\
   & SVM & 0.287 & 0.313 & 0.322 & 0.313 & 0.361 & 0.388 & & &  \\   \hline
  \multirow{5}*{\emph{riTDS-1}}& J48 & 0.337 & 0.347 & 0.371 & 0.400 & 0.430 & 0.466 &\multirow{5}*{316.7} & \multirow{5}*{537.9} & \multirow{5}*{722.9} \\
   & LR  & 0.334 & 0.365 & 0.362 & 0.393 & 0.445 & 0.448 & & & \\
   & NB  & \textbf{0.437} & \textbf{0.461} & \textbf{0.465} & \textbf{0.565} & \textbf{0.595}  & \textbf{0.606} & & & \\
   & RF  & 0.327 & 0.308 & 0.309 & 0.388 & 0.392 & 0.405 & & & \\
   & SVM & 0.292 & 0.320 & 0.319 & 0.315 & 0.368 & 0.385 & & & \\   \hline
  \multirow{5}*{\emph{riTDS-2}}& J48 & 0.325 & 0.307 & 0.340 & 0.383 & 0.396 & 0.429 &\multirow{5}*{218.7} & \multirow{5}*{286.7} & \multirow{5}*{317.7} \\
   & LR  & 0.344 & 0.359 & 0.369 & 0.417 & 0.448 & 0.452 & & & \\
   & NB  & \textbf{0.453} & \textbf{0.464} &\textbf{ 0.475} & \textbf{0.585} & \textbf{0.599} & \textbf{0.613} & & & \\
   & RF  & 0.311 & 0.315 & 0.327 & 0.361 & 0.401 & 0.422 & & & \\
   & SVM & 0.287 & 0.312 & 0.310 & 0.306 & 0.365 & 0.381 & & & \\   \hline
  \multirow{5}*{\emph{iTDS}}& J48 & 0.340 & 0.338 & 0.343 & 0.469 & 0.440 & 0.467 &\multirow{5}*{209.8} & \multirow{5}*{503.4} & \multirow{5}*{697.2} \\
   & LR  & 0.357 & 0.346 & 0.338 & 0.477 & 0.450 & 0.442 & & &  \\
   & NB  & \textbf{0.466} & \textbf{0.460} & \textbf{0.458} & \textbf{0.611} & \textbf{0.610} & \textbf{0.610} & & &  \\
   & RF  & 0.336 & 0.319 & 0.324 & 0.452 & 0.427 & 0.441 & & &  \\
   & SVM & 0.310 & 0.300 & 0.305 & 0.394 & 0.381 & 0.389 & & & \\
  \hline
\end{tabular}
\end{table*}

In fact, the difference between the training set-driven filter and the test set-driven filter is determined by which data set (TDS or test) contains more information about defects \citep{Peters:Better}. To reflect the comparison of defect information between TDS and the test set, the concept of defect proneness ratio (\emph{DPR}) is introduced in our experiments. \emph{DPR} represents the ratio of the proportion of defects in the training set to the proportion of defects in the test set. Intuitively, when the value is approximately one, the relative proportions of defects in TDS and in the test set reach equilibrium.
      \begin{equation}\label{DPR}
        DPR=\frac{\%Defects(training set)}{\%Defects(test set)}.
      \end{equation}

%
%
\subsection{Results}
We organize our results according to the three research questions proposed in Section 1.

\emph{RQ1: Does our TDS simplification method perform well compared with the benchmark methods?}

Given the strategies for TDS simplification at different levels of granularity, Table \ref{Tab.RQ-1} shows some interesting results. First, the fine-grained strategy (\emph{iTDS}) outperforms the coarse-grained strategy (\emph{rTDS}) as a whole, indicated by the greater mean values of evaluation measures, especially for the \emph{g-measure}. For example, the \emph{g-measure} mean values of the \emph{rTDS} with Na\"{\i}ve Bayes are 0.552, 0.592 and 0.594, respectively, but they are 0.611, 0.610 and 0.610 for the \emph{iTDS}, respectively. Second, the result of the \emph{riTDS} is approximately on the borderline between the \emph{rTDS} and the \emph{iTDS}, as it is a combination of the two methods, whereas some \emph{f-measure} mean values of the \emph{riTDS} are even better for those prediction models built with Logistic Regression and Na\"{\i}ve Bayes. Third, three out of five predictors (i.e., those built with LR, NB and RF) present a better \emph{f-measure} and \emph{g-measure} mean values with the \emph{riTDS-2}, in particular, when increasing the value of the parameter $r$. That is, the filter based on training set-driven filtering strategy may in general work better on the instance-level simplification. It is worthwhile to note that the value of the parameter \emph{k} mentioned in Algorithm \ref{Algorithm1} (line 15 and 21) is set to 10 because the same assignment was used in the prior studies \citep{Peters:Better,Turhan:On}.

Regarding the necessity of TDS simplification, we then investigated the size of the final simplified TDS actually used to train defect predictors. As shown in Table \ref{Tab.RQ-1}, the last three columns list the corresponding average number of instances in simplified TDS in each scenario. Although the effect of the \emph{riTDS} method on prediction is not always distinct, it is more effective from the perspective of TDS simplification. More specifically, compared with the simplification at a single level of granularity, there is a several-fold decease in the number of useless instances with an increase of $r$, especially for the \emph{riTDS-2}. Furthermore, it is obvious from Table \ref{Tab.RQ-1} that a large increase in TDS's size (e.g., from 317 to 1222) does not significantly improve prediction performance, and sometimes it is just the opposite. That is, to a certain extent, the quality rather than the quantity of training data is a crucial factor that affects the performance of CPDP. This is one of our primary motivations to simplify the training data in this study.

\begin{table*}\small
\centering
\caption{ A comparison between \emph{riTDS} and \emph{iTDS}. riTDS/iTDS represents the ratio of the mean of the former to that of the latter, and riTDS vs. iTDS means the Wilcoxon signed-rank test of the distribution of prediction results of the two methods in terms of \emph{f-measure} and \emph{g-measure}.}\label{test1}
\begin{tabular}{c|c|ccc|ccc|ccc|ccc}
  \hline
  \multicolumn{2}{c|}{\multirow{3}*{ Methods}} & \multicolumn{6}{c|}{\emph{f-measure}} & \multicolumn{6}{c}{\emph{g-measure}} \\
  \cline{3-14}
  \multicolumn{2}{c|}{}  & \multicolumn{3}{c|}{riTDS/iTDS} & \multicolumn{3}{c|}{vs. iTDS ($Sig. p=0.01$)} &\multicolumn{3}{c|}{riTDS/iTDS} & \multicolumn{3}{c}{vs. iTDS ($Sig. p=0.01$)} \\
  \cline{3-14}
  \multicolumn{2}{c|}{} & 1 & 2 & 3 & 1 & 2 & 3 & 1 & 2 & 3 & 1 & 2 & 3
   \\ \hline
  \multirow{5}*{\emph{riTDS-1}} & J48 & 0.991 & 1.029 & 1.084 & 0.700 & 0.884 & 0.270 & 0.852 & 0.978 & 0.997 & 0.228 & 0.980 & 0.739\\
   & LR  &  0.936 & 1.055 & 1.071 & 0.871 & 0.489 & 0.469 & 0.824 & 0.989 & 1.014 & 0.158 & 0.858 & 0.782\\
   & NB  &  0.937 & 1.002 & 1.015 & 0.086 & 0.765 & 0.549 & 0.924 & 0.974 & 0.994 & 0.012 & 0.437 & 0.993\\
   & RF  &  0.974 & 0.965 & 0.953 & 0.791 & 0.437 & 0.782 & 0.858 & 0.918 & 0.917 & 0.096 & 0.164 & 0.544\\
   & SVM &  0.943 & 1.067 & 1.046 & 0.533 & 0.871 & 0.844 & 0.801 & 0.965 & 0.988 & 0.144 & 0.752 & 0.966\\   \hline
  \multirow{5}*{\emph{riTDS-2}} & J48 & 0.954 & 0.910 & 0.994 & 0.980 & 0.626 & 0.858 & 0.817 & 0.899 & 0.919 & 0.139 & 0.533 & 0.578 \\
   & LR  & 0.964 & 1.039 & 1.092 & 0.859 & 0.544 & 0.369 & 0.874 & 0.997 & 1.023 & 0.489 & 0.651 & 0.688\\
   & NB  & 0.973 & 1.008 & 1.037 & 0.343 & 0.858 & 0.285 & 0.956 & 0.981 & 1.006 & 0.203 & 0.293 & 0.437\\
   & RF  & 0.998 & 0.989 & 1.006 & 0.457 & 0.884 & 0.726 & 0.799 & 0.938 & 0.958 & 0.027 & 0.285 & 0.925\\
   & SVM & 0.925 & 1.041 & 1.016 & 0.427 & 0.966 & 0.912 & 0.776 & 0.956 & 0.977 & 0.080 & 0.925 & 0.993\\
  \hline
\end{tabular}
\end{table*}

\begin{table*}\small
\centering
\caption{ A comparison of the precision of the \emph{riTDS} and the \emph{iTDS}, and $\Delta$ represents the relative increment of precision.}\label{precision}
\begin{tabular}{c|c|ccc|ccc}
  \hline
  \multicolumn{2}{c|}{\multirow{2}*{ Methods}}  & \multicolumn{3}{c|}{precision} &\multicolumn{3}{c}{$\Delta$ (riTDS-iTDS)} \\
  \cline{3-8}
  \multicolumn{2}{c|}{} & 1 & 2 & 3 & 1 & 2 & 3
   \\ \hline
  \multirow{5}*{\emph{riTDS-1}} & J48 & 0.437 & 0.403 & 0.426 & 0.001 & -0.054 & 0.024 \\
   & LR  &  0.435 & 0.432 & 0.413 & 0.044 & 0.054 & 0.066 \\
   & NB  &  0.550 & 0.556 & 0.560 & 0.030 & 0.030 & 0.066 \\
   & RF  &  0.398 & 0.334 & 0.345 & 0.003 & 0.003 & 0.040 \\
   & SVM &  0.391 & 0.366 & 0.337 & -0.002 & 0.004 & 0.009 \\   \hline
  \multirow{5}*{\emph{riTDS-2}} & J48 & 0.427 & 0.360 & 0.405 & 0.066 & 0.047 & 0.051 \\
   & LR  & 0.472 & 0.449 & 0.429 & 0.110 & 0.123 & 0.106 \\
   & NB  & 0.576 & 0.584 & 0.612 & -0.002 & 0.028 & 0.057 \\
   & RF  & 0.381 & 0.352 & 0.358 & 0.058 & 0.050 & 0.043 \\
   & SVM & 0.380 & 0.358 & 0.347 & 0.108 & 0.094 & 0.077 \\
  \hline
\end{tabular}
\end{table*}

\begin{figure*}
\centering
\includegraphics[width=5in,height=2.2in]{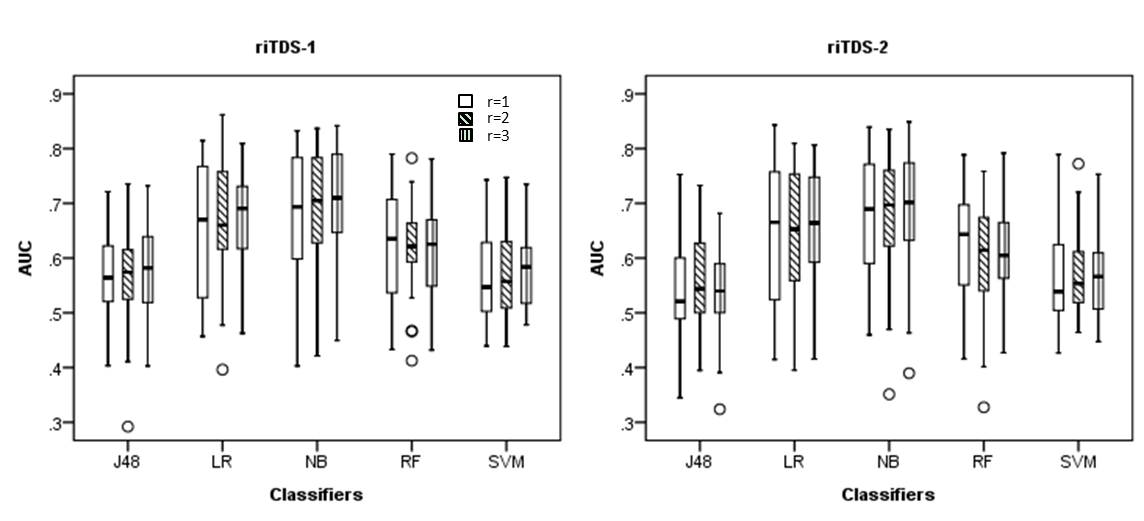}
\caption{The standardized boxplots of the distributions of AUC values based on the \emph{riTDS-1} and the \emph{riTDS-2}. From the bottom \\to the top of a standardized box plot: minimum, first quartile, median, third quartile and maximum. The outliers are plotted as circles.}
\label{auc}
\end{figure*}

\begin{figure*}
\centering
\includegraphics[width=7in,height=2.5in]{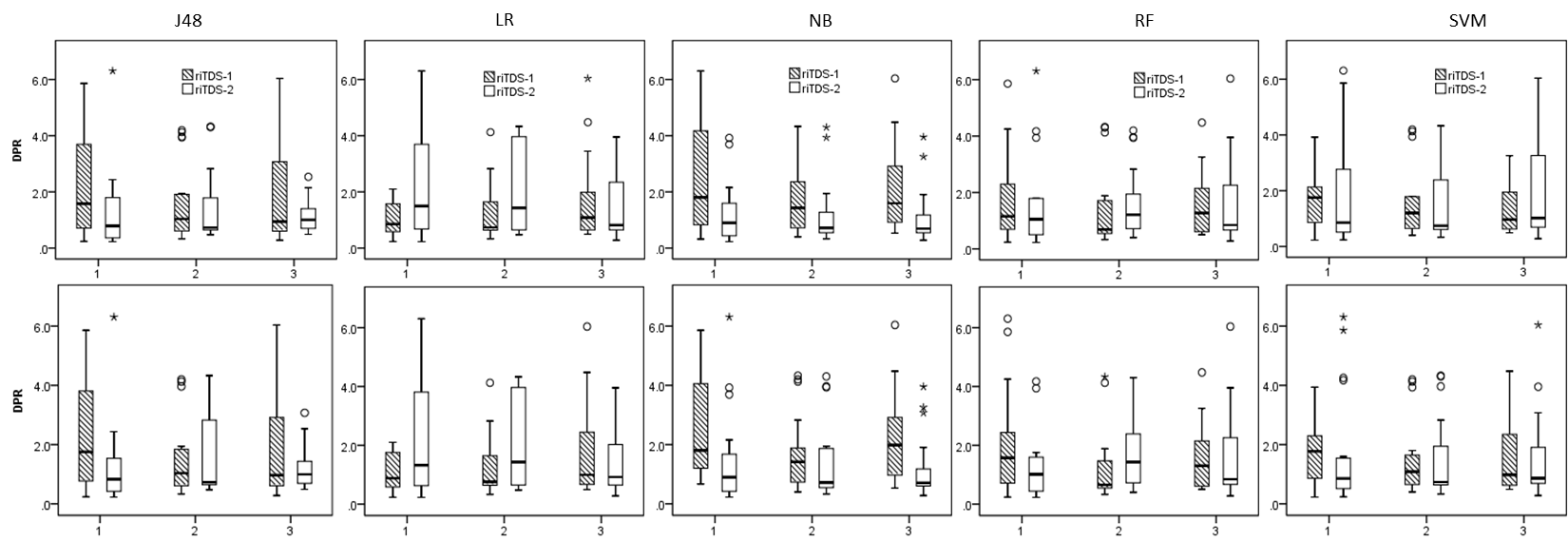}
\caption{The standardized boxplots of the DPR distribution of predictions in the groups riTDS-1 and riTDS-2 using \emph{f-measure} (up) and \emph{g-measure} (down) as the group division standard. From the bottom to the top of a standardized box plot: minimum, first quartile, median, third quartile and maximum.The outliers are plotted as circles and pentagrams.}
\label{Fig.RQ3-2}
\end{figure*}

\begin{figure*}
\centering
\includegraphics[width=7in,height=2.5in]{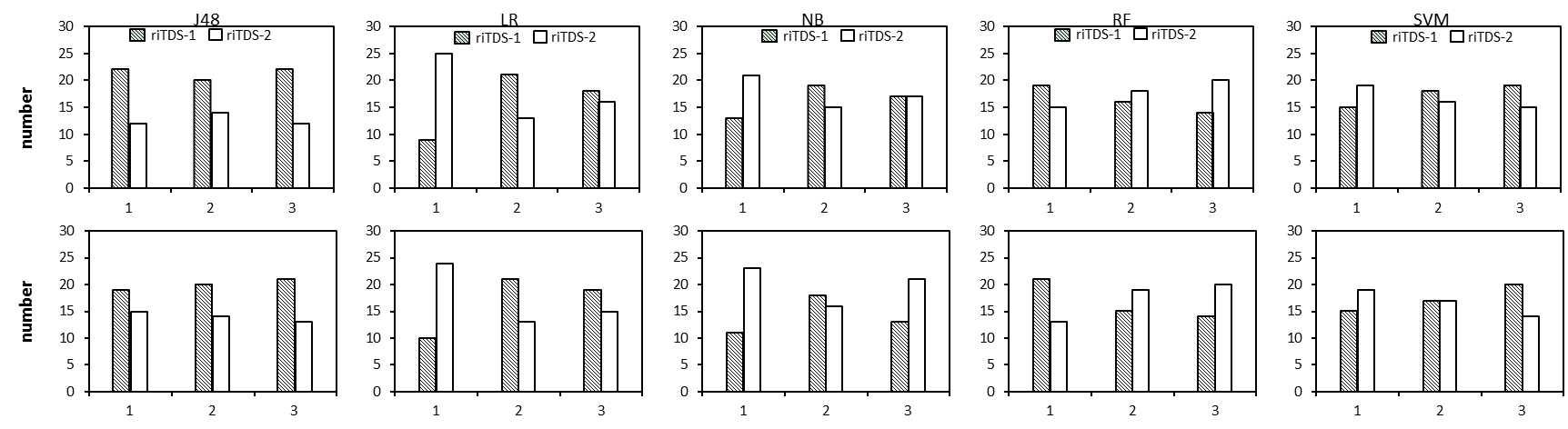}
\caption{The comparison between the groups riTDS-1 and riTDS-2 using \emph{f-measure} (up) and \emph{g-measure} (down) as the group division standard. The number of elements in the groups is counted among the 34 CPDP cases.}
\label{Fig.RQ3-1}
\end{figure*}

To further investigate the practicability of our TDS simplification method, we compared the performance of the \emph{riTDS} with the \emph{iTDS} from the viewpoint of statistically significant difference. Table \ref{test1} presents the results of the Wilcoxon signed-rank test based on the \emph{null hypothesis} that the medians of the two methods are identical (i.e., $H_0: \mu_1=\mu_2$). Obviously, the results highlight that there is no significant difference between the \emph{riTDS} and the \emph{iTDS}, indicated by all of the $p>0.05$ cases for the five typical classifiers. In other words, this suggests that the \emph{riTDS} method can achieve satisfactory performance under the premise of using fewer instances for training, compared with the benchmark method.

Moreover, the \emph{riTDS} method with different filters can achieve better precision than the \emph{iTDS} method. In Table \ref{precision}, it is clear that the degree of precision improvement of the \emph{riTDS-2} is greater than that of the \emph{riTDS-1}, and the results of the \emph{riTDS-2} with LR and SVM are more significant. Therefore, our TDS simplification method not only achieved a comparative \emph{f-measure} and \emph{g-measure} values, but also significantly reduced the number of training instances and improved the performance in terms of precision.

\emph{RQ2: Which classifier is more suitable for CPDP with our TDS simplification method?}

The numbers in bold in Table \ref{Tab.RQ-1} indicate that the predictor built with Na\"{\i}ve Bayes yields the best performance because of the greatest \emph{f-measure} and \emph{g-measure} mean values, followed by those built with Logistic Regression and J48. With regard to AUC value, Figure \ref{auc} further validates that Na\"{\i}ve Bayes is the best classifier and that Logistic Regression is an alternative in our context. However, J48 presents an obvious disadvantage because of its lower median AUC value, although it shows middling performance in terms of \emph{f-measure} and \emph{g-measure} mean values.

Interestingly, whichever level of granularity we select, the predictor built with SVM seems to have the worst performance, especially when using the \emph{rTDS} method. Our results also validate the statement that simple learning algorithms tend to perform well for defect prediction \citep{Hall:A}. In the literature \citep{Herbold:Training}, the author weighted the training instances, thus leading to a remarkable performance improvement by the SVM classifier. The reason why we did not take the weight of training data into account is that we focused primarily on understanding the differences between TDS simplification methods from the perspective of granularity (e.g., release-level vs. instance-level). Hence, we used the same data processing method for all classifiers under discussion, without considering specific optimization for any one of the classifiers.

In addition, for each scenario ($r=1,2,3$), we divided the 34 CPDP cases into two groups according to their performance measures (\emph{f-measure} and \emph{g-measure}). That is, for the $i^{th}$ target release, if $measure_{ir}^{riTDS-1}>measure_{ir}^{riTDS-2}$, this CPDP is classified into the group riTDS-1; otherwise, it belongs to the group riTDS-2. We then compared the distribution of \emph{DPR} values between the group riTDS-1 and the group riTDS-2 in terms of \emph{f-measure} and \emph{g-measure}.  Figure \ref{Fig.RQ3-2} shows that for those predictions with Na\"{\i}ve Bayes, the group riTDS-1 has a significantly higher median \emph{DPR} value than the group riTDS-2, and this trend is independent of the parameter $r$. Specifically, the median DPR values of the former are more than twice those of the latter. For example, the median \emph{DPR} values of the two groups are 1.59 vs. 0.691 (\emph{f-measure}) and 1.98 vs. 0.706 (\emph{g-measure}), respectively, when returning the top three releases as the set rTDS. In addition, J48 and SVM show a similar trend except in the scenario $r=3$. The obvious difference in \emph{DPR} values of CPDP is a meaningful insight into how to determine an appropriate filter for instance simplification in the \emph{riTDS}. Therefore, the predictor built with Na\"{\i}ve Bayes is still the most suitable prediction model due to its ability to distinguish different filters. The discussion on filter selection in terms of \emph{DPR} will be introduced in the following subsection.

%
%

\begin{figure*}
\centering
\includegraphics[width=7in,height=2.5in]{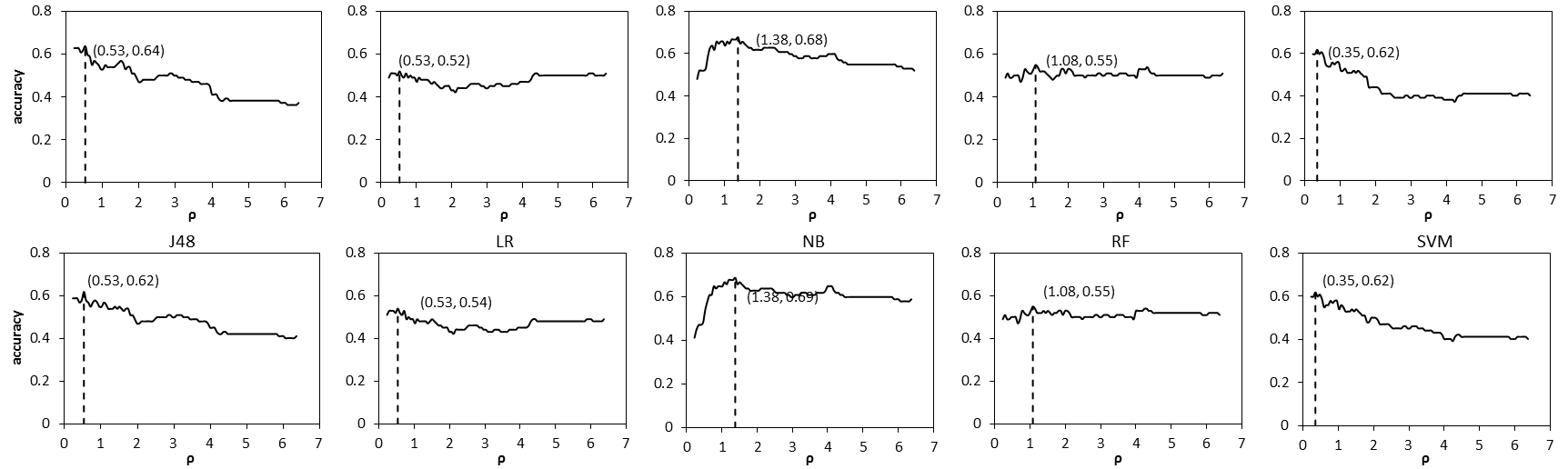}
\caption{The recommendation $accuracy$ value changes with the threshold $\rho$ of \emph{DPR} values according to the assumption that the \emph{riTDS-1} filter is recommended if $DPR\geqslant \rho$ (named as $\rho +$). The groupings in Figure \ref{Fig.RQ3-1} are viewed as the actual results in our experiment. }
\label{Fig.RQ3-3-1}
\end{figure*}

\begin{figure*}
\centering
\includegraphics[width=7in,height=2.5in]{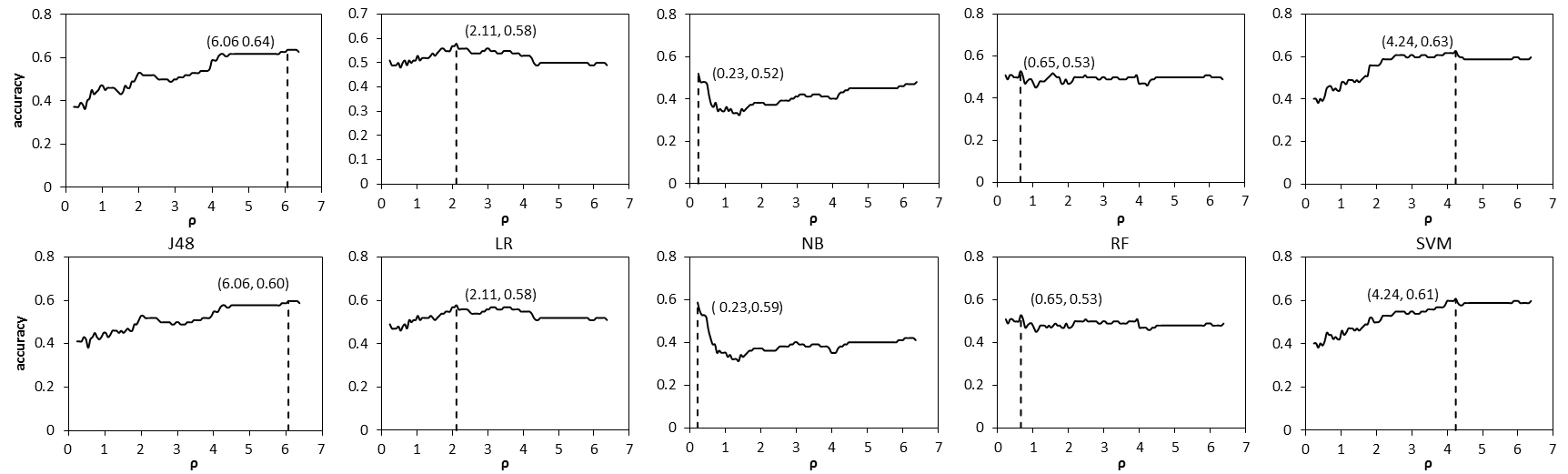}
\caption{The recommendation $accuracy$ value changes with the threshold $\rho$ of \emph{DPR} values according to the opposite assumption that the \emph{riTDS-2} filter is recommended if $DPR\geqslant \rho$ (named as $\rho -$). The groupings in Figure \ref{Fig.RQ3-1} are viewed as the actual results in our experiment.}
\label{Fig.RQ3-3-2}
\end{figure*}

\begin{figure*}
\centering
\includegraphics[width=5in,height=2in]{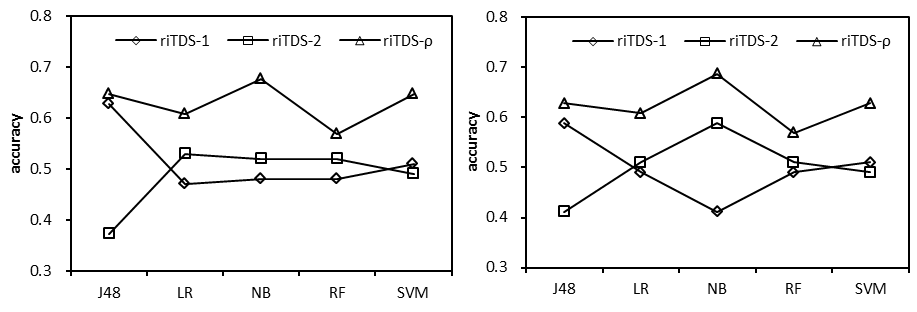}
\caption{The comparison of recommendation $accuracy$ among the predictors built with different filters using \emph{f-measure} (left) and \emph{g-measure} (right) as the group division standard. }
\label{Fig.RQ3-4}
\end{figure*}

\emph{RQ3: Which filter for TDS simplification should be preferable in a specific scenario?}

For different scenarios about $r$, on the basis of the aforementioned groups, Figure \ref{Fig.RQ3-1} shows the number of elements in each group. Although the results of LR and SVM are similar to each other, there are no universal patterns for all classifiers. The results indicate that some CPDP cases are indeed preferable to the \emph{riTDS-1}, while others are yet apt to use the \emph{riTDS-2}. For example, for all scenarios with J48, the group riTDS-1 has higher bars than the group riTDS-2 using both \emph{f-measure} and \emph{g-measure} as the group division standard, which, in turn, has more elements when using Random Forest except in the scenario $r=1$. Thus, it is very clear that the above findings drawn from Figure \ref{Fig.RQ3-2} and Figure\ref{Fig.RQ3-1} only show an overall difference between the two filters for instance simplification in CPDP (\emph{riTDS-1} and \emph{riTDS-2}), but they cannot yet effectively help us make a reasonable decision on the choice of an appropriate filtering strategy.

To solve this problem, we first gathered the 102 ($3\times 34 = 102$) predictions used in our experiments, and then divided them into two groups according to the similar rule mentioned above. The groups riTDS-1 and riTDS-2 will be viewed as the actual observations in the following tasks. According to the \emph{DPR} distribution in Figure \ref{Fig.RQ3-2}, we suppose that the \emph{riTDS-1} filter is recommended to a target release if its \emph{DPR} value is not less than $\rho$; otherwise, the \emph{riTDS-2} filter is recommended. This assumption is named as $\rho +$. Thus, the value of \emph{accuracy} is calculated using the Eq. (\ref{accuracy}), where $TP$ and $TN$ represent the correct recommendation for the groups riTDS-1 and riTDS-2 with a specific $\rho$, respectively. Note that, $\rho \in [min, max]$, where $min$ and $max$ are the minimum and maximum \emph{DPR} values among the 102 predictions. The higher the \emph{accuracy} value is, the more reliable the choice of filters made by $\rho$. Figure \ref{Fig.RQ3-3-1} shows that the \emph{accuracy} values reach a peak when $\rho$ changes from the minimum to the maximum. With the optimal \emph{accuracy} value, it is not hard to make a choice between the \emph{riTDS-1} filter and the \emph{riTDS-2} filter when using a specific classifier. That is, we can employ the parameter $\rho$ as a corresponding threshold to determine the eventual choice of filtering strategies. Interestingly, each classifier has the same optimal $\rho$ value using whichever measure as the group division standard. For example, with respect to Na\"{\i}ve Bayes, the \emph{riTDS-2} filter should be recommended if the \emph{DPR} value of a target release is 1.0; otherwise, the \emph{riTDS-1} filter should be preferable if the value equals 1.5.

To further identify the appropriate threshold of the $\rho$ value for each classifier, we conducted another experiment with the opposite assumption (named as $\rho -$). That is, the \emph{riTDS-2} filter is recommended to a target release if its \emph{DPR} value is not less than $\rho$; otherwise, the \emph{riTDS-1} filter is recommended. In Figure \ref{Fig.RQ3-3-2}, the overall optimal \emph{accuracy} values of four cases declined, in particular for the case of Na\"{\i}ve Bayes where the maximum values are only 0.52 and 0.59 when using \emph{f-measure} and \emph{g-measure} as the group division standard, respectively. In fact, these two results indicate the case in which all predictions used the \emph{riTDS-2} filter because 0.23 is the lowest \emph{DPR} value. However, Logistic Regression achieves a higher \emph{accuracy} and larger optimal $\rho$ value according to the opposite assumption, which is consistent with the \emph{DPR} distribution shown in Figure \ref{Fig.RQ3-2}. According to the results of Figure \ref{Fig.RQ3-3-1} and Figure \ref{Fig.RQ3-3-2}, the threshold of $\rho$ used to determine the \emph{riTDS-1} filter for each classifier can be identified in Table \ref{range}, whereas the corresponding complementary set is suitable for the \emph{riTDS-2} filter.

\begin{table}\small
\centering
\caption{ The threshold of $\rho$ (the range of DPR) for the \emph{riTDS-1} filter.}\label{range}
\begin{tabular}{c|c}
  \hline
  \multicolumn{1}{c|}{Classifier}  & \multicolumn{1}{c}{Range}  \\
   \hline
    J48 & $0.53 \leqslant DPR < 6.06 $ \\
    LR  & $0.53 \leqslant DPR < 2.11 $  \\
    NB  &  $1.38 \leqslant DPR \quad or \quad DPR<0.23 $ \\
    RF  &  $1.08 \leqslant DPR \quad or \quad DPR<0.65 $ \\
    SVM &  $0.35 \leqslant DPR < 4.24 $  \\   \hline
\end{tabular}
\end{table}

With the threshold of $\rho$, we compared the recommendation $accuracy$ among the three cases with different filters: the \emph{riTDS-1} filter, the \emph{riTDS-2} filter, and filter selection determined by the \emph{DPR} value. The results show that our approach increases the \emph{accuracy} value, in particular for Logistical Regression, Na\"{\i}ve Bayes and SVM (see Figure \ref{Fig.RQ3-4}). For example, compared with the filters \emph{riTDS-1} and \emph{riTDS-2}, for Na\"{\i}ve Bayes, the $riTDS-\rho$ filter achieves a marked increase in $accuracy$ when using the \emph{f-measure} and \emph{g-measure} as the group division standard. The values in terms of different groupings grow by $40.8\%$ and $30.2\%$, and $66.7\%$ and $16.7\%$, respectively (see Table \ref{increment}). The improvement of recommendation $accuracy$ indicates that our approach to determining the appropriate filter for TDS simplification is feasible and outperforms the \emph{riTDS} with a single type of filter.

\begin{table}\small
\centering
\caption{ The comparison between different filters with regard to recommendation $accuracy$.}\label{increment}
\begin{tabular}{c|c|ccccc}
  \hline
  Grouping  & riTDS & J48 & LR & NB & RF & SVM \\
   \hline
    \multicolumn{1}{c|}{\multirow{3}*{\emph{f-measure}}} & -$\rho$ & 0.647 & 0.608 & 0.676 & 0.569 & 0.647 \\
     & -1 (\%) & +3.1 & +29.2 & +40.8 & +18.4 &+26.9 \\
     & -2 (\%) & +73.7 & +14.8 & +30.2 & +9.4 & +32.0 \\ \hline
    \multicolumn{1}{c|}{\multirow{3}*{\emph{g-measure}}} & -$\rho$ & 0.627 & 0.608 & 0.686 & 0.569 & 0.627 \\
     & -1 (\%) & +6.7 & +24.0 & +66.7 & +16.0 & +23.1 \\
     & -2 (\%) & +52.4 & +19.2 & +16.7 & +11.5 & +28.0 \\ \hline
\end{tabular}
\end{table}

\begin{figure*}
\centering
\includegraphics[width=7in,height=2.5in]{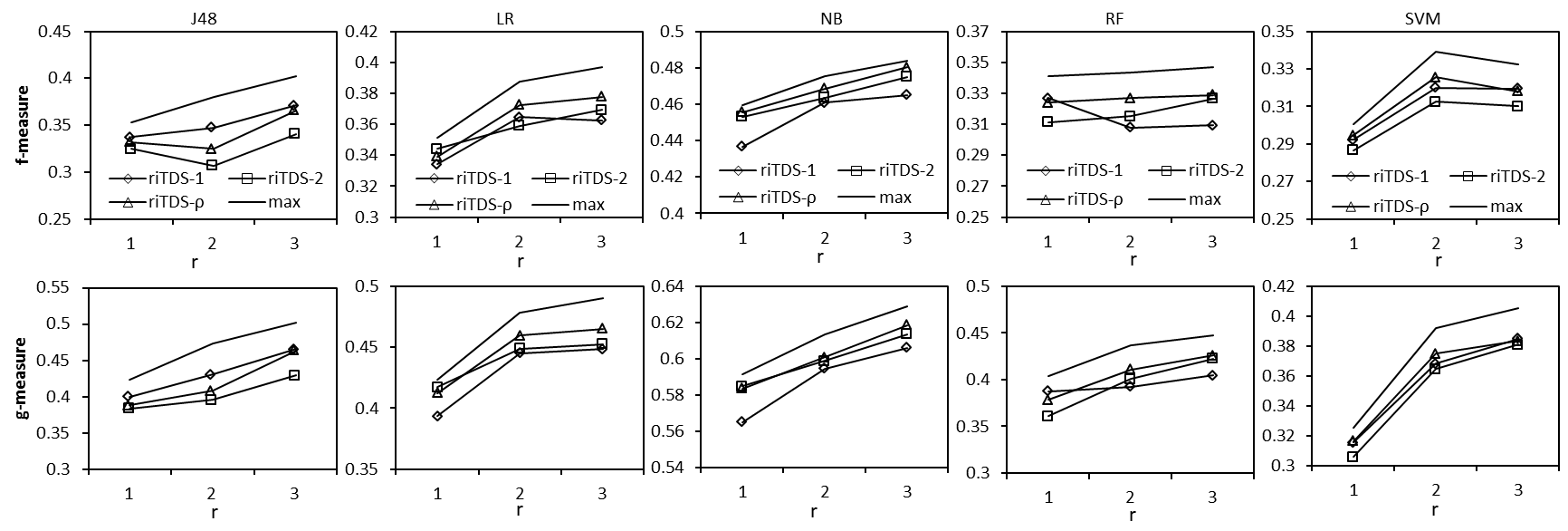}
\caption{The improvement of \emph{f-measure} and \emph{g-measure} for $riTDS-\rho$.}
\label{Fig.RQ4}
\end{figure*}

To further validate the feasibility of our approach, we compared the prediction performance among the three cases in terms of \emph{f-measure} and \emph{g-measure}. Figure \ref{Fig.RQ4} shows that our approach achieves various degrees of improvement in the \emph{f-measure} and \emph{g-measure} values overall, in particular for the prediction built with Logistical Regression and Na\"{\i}ve Bayes when r is 2 and 3. Note that, the improvement is optimistic compared with the best case where the filter with the greater measure value is applied to a target release, although the degree of improvement does not seem to be great. In addition, we also compared the prediction $precision$ of our approach with the two benchmark filters. Again, there is an overall growth trend for the five classifiers (see Table \ref{incr_precision}). This evidence suggests that our approach is also feasible in terms of prediction performance.

\begin{table}\small
\centering
\caption{ The increment of prediction \emph{precision} for $riTDS-\rho$ .}\label{incr_precision}
\begin{tabular}{c|c|ccc}
  \hline
   Classifier  & riTDS & 1 & 2 & 3 \\
   \hline
    \multicolumn{1}{c|}{\multirow{3}*{J48}} & -$\rho$ & 0.437 & 0.395 & 0.428 \\
     & -1  & 0.000  & -0.008 & +0.002  \\
     & -2  & +0.010 & +0.035 & +0.023  \\ \hline
     \multicolumn{1}{c|}{\multirow{3}*{LR}} & -$\rho$ & 0.446 & 0.444 & 0.420 \\
     & -1  & +0.011  & +0.012 & +0.007  \\
     & -2  & -0.026 & -0.005 & -0.009  \\ \hline
     \multicolumn{1}{c|}{\multirow{3}*{NB}} & -$\rho$ & 0.577 & 0.575 & 0.612 \\
     & -1  & +0.027  & +0.019 & +0.052  \\
     & -2  & +0.001 & -0.009 & 0.000  \\ \hline
     \multicolumn{1}{c|}{\multirow{3}*{RF}} & -$\rho$ & 0.396 & 0.339 & 0.364 \\
     & -1  & -0.002 & +0.005 & +0.019  \\
     & -2  & +0.015 & -0.013 & +0.006  \\ \hline
     \multicolumn{1}{c|}{\multirow{3}*{SVM}} & -$\rho$ & 0.391 & 0.357 & 0.348 \\
     & -1  & 0.000  & -0.009 & +0.011  \\
     & -2  & +0.011 & -0.001 & +0.001  \\ \hline
\end{tabular}
\end{table}

\section{Discussion}

\emph{RQ1:} A larger amount of training data may not lead to a higher performance of CPDP, suggesting the necessity of simplifying training data. However, none of the existing methods take the levels of granularity of data into consideration, especially with regard to multiple granularity (e.g., the two-step strategy proposed in our paper), which is a key factor for building practical CPDP models. As we consider such a factor, our experimental results show that less instances are involved in training the predictors based on multiple levels of granularity compared with those based on a single level of granularity, with little loss of accuracy. The simplified TDS preserves the most relevant training instances, which is helpful to reduce the false alarms and build the quality predictors.

The prediction results of different predictors based on the methods \emph{rTDS}, \emph{iTDS} and \emph{riTDS} were calculated without any feature selection techniques. That is, for the simplified TDS, all predictors were built with the twenty software metrics (viz. features). As shown in Figure \ref{summary}, this paper focuses on how to reduce data volumes in a TDS. If we applied feature selection techniques to building defect predictors, it is hard to distinguish what factor actually obtained the greater improvement on prediction performance. Therefore, we did not consider feature selection in our experiments. Additionally, the parameter $r$ was set to no more than 3 because 8 out of 10 projects under discussion have no more than 4 releases available. That is, the majority of projects have to select no more than 3 releases as training data even if we conduct experiments on WPDP. Prior studies \citep{He:An} have also used the same setting for the parameter $r$.

\emph{RQ2:} As we know, Na\"{\i}ve Bayes has been validated as a robust machine learning algorithm for supervised software defect prediction problems in both WPDP and CPDP. Interestingly, our result is completely consistent with the conclusions drawn in the literature \citep{Hall:A,Catal:Software}, that is, Na\"{\i}ve Bayes outperforms the other typical classifiers within our CPDP context in terms of \emph{f-measure}, \emph{g-measure} and AUC. It is worthwhile to note that different prediction models were built based on these classifiers without specific optimization because in this study, we focused primarily on the levels of granularity and filtering strategies for TDS simplification. However, the performance differences between different prediction models indicate that simple classifiers, such as Na\"{\i}ve Bayes, can be preferable to training data of quality.

In addition, for \emph{DPR}, different classifiers exhibit different abilities to distinguish the results of the two filters in question. For example, the group riTDS-1 has a higher median \emph{DPR} value than the group riTDS-2 except Logistical Regression when the parameter $r$ is 1. However, the opposite results occur using Logistical Regression and Random Forest when $r$ is 2. J48 and SVM have the approximate median \emph{DPR} values between the group riTDS-1 and the group riTDS-2 when $r$ is 3, although they maintain a similar trend in the first two $r$  values. However, the best ability of Na\"{\i}ve Bayes to distinguish the group riTDS-1 and the group riTDS-2 paves a way for the feasibility and generality of our approach proposed to answer \emph{RQ3}.

\emph{RQ3:} As an alternative strategy, the training set-driven filter for TDS simplification is in general better than the test set-driven filter, which is consistent with the findings obtained in \citep{Peters:Better}. However, the authors did not analyze the specific application scenarios for each type of filter. We filled the gap in terms of recommendation $accuracy$ based on \emph{DPR} value, and found that the training set-driven filter is more suitable for those predictions with very low or very large \emph{DPR} values when using J48, LR, and SVM classifiers. Conversely, a prediction with a middle \emph{DPR} value is more likely to choose the training set-driven filter when using NB and RF classifiers. Note that, to make the right decision between the training set-driven filter and the test set-driven filter according to the value of \emph{DPR}, we seek the optimal point $\rho$ through gradually changing the value of \emph{DPR} with an increment of $\frac{max-min}{100}$.

With regard to the threshold of $\rho$, we have to admit that we may obtain different thresholds for such an index if other formulas are used to evaluate the recommendation results. Nevertheless, we still obtained various valuable findings. For example, the test set-driven filter is preferable when the \emph{DPR} value is between 1.38 and 2.11, and this range is suitable for all five of the classifiers in our context. Although there are no common ranges for training set-driven filter selection, our results still indicate that the practical guideline for the decision-making on which filtering strategy is suitable for instance selection indeed exists , and that it does improve the prediction performance of those predictors based on a single type of filter.

\section{Threats to Validity}
In this study, although we obtained several interesting findings according to the three research questions proposed in Section 1, some potential threats to the validity of our work still exist.

Threats to \emph{construct validity} are primarily related to the data sets we used. All of the data sets were collected by Jureczko and Madeyski \citep{Jureczko:Towards} and Jureczko and Spinellis \citep{Jureczko:Using} with the support of existing tools: BugInfo and Ckjm. These data sets have been validated and applied to several prior studies, though errors in the process of defect identification may exist. Therefore, we believe that our results are credible and can be reproduced. Additionally, we applied a log transformation to feature values before building defect predictors, and we cannot ensure that it is better than other preprocessing methods. The impact of data preprocessing on prediction performance is also an interesting problem that needs further investigation.

Threats to \emph{internal validity} are mainly related to various learning algorithm settings in our study. For our experiments, although the $k$-nearest neighbors algorithm (KNN) was selected as the basic selection algorithm, we are aware that our results would change if we were to use a different method. However, to the best of our knowledge, both KNN and its variants were successfully applied to TDS simplification in several prior studies \citep{Peters:Better,Herbold:Training}. Moreover, we did not implement specific optimization for any classifiers in question when building different prediction models because the goal of this experiment is not to improve the performance of a given classifier.

Threats to \emph{external validity} could be related to the generality of the results to other on-line public data sets used for defect prediction, such as NASA and Mozilla. The data sets used in our experiments are chosen from a small subset of all projects in the PROMISE repository, and it is possible that we accidentally selected data sets that have better (or worse) than average CPDP performance, implying that some of our findings (e.g., the threshold of $\rho$ for the five typical classifiers) might not be generalizable to other data sets.

\section{Conclusion}
TDS simplification, which filters out the irrelevant and redundant training data, plays an important role in building better CPDP models. This study reports an empirical study aiming at investigating the impact of the level of granularity and filtering strategy on TDS simplification. The study has been conducted on 34 releases of 10 open-source projects in the PROMISE repository and consists of (1) a comparison between multi-granularity and benchmark (single level of granularity) TDS simplification, (2) a selection of the best classifier in our context, and (3) an assessment of practical selection rules for the state-of-the-art filtering strategies for instance simplification.

The results indicate that the CPDP predictions based on the multi-granularity simplification approach (e.g., the two-step strategy proposed in our paper) capture competitive \emph{f-measure} and \emph{g-measure} values showing no statistically significant differences compared with those benchmark TDS simplification approaches, and that the size of simplified TDS was sharply reduced with an increase in the number of returned neighbors at the level of release. In addition, our results also show that more actually defective instances can be predicted by our method and that Na\"{\i}ve Bayes is more suitable for building predictors for CPDP with simplified TDS. Finally, the \emph{DPR} index is useful in determining a proper filtering strategy when using the \emph{riTDS} method, and the practical selection rule based on the \emph{DPR} value does improve prediction performance to some extent.

Our future work will focus mainly on two aspects: on the one hand, we will collect more open-source projects (e.g., Eclipse and Mozilla) to validate the generality of our approach; on the other hand, we will further consider the number of defects of an instance to provide an effective TDS simplification method for CPDP.


\section*{Acknowledgment}
This work is supported by the National Basic Research Program of China (No. 2014CB340401), the National Natural Science Foundation of China (Nos. 61273216, 61272111, 61202048 and 61202032), the Science and Technology Innovation Program of Hubei Province (No. 2013AAA020), the National Science and Technology Pillar Program of China (No. 2012BAH07B01), the open foundation of Hubei Provincial Key Laboratory of Intelligent Information Processing and Real-time Industrial System (No. znss2013B017), and the Youth Chenguang Project of Science and Technology of Wuhan City in China (No. 2014070404010232).


\end{document}